\documentclass[doublecol]{epl2} 
\usepackage{graphicx}
\usepackage{amssymb}
\usepackage{wasysym}

\newcommand{\rt}{\rightarrow}
\def\be{\begin{equation}}
\def\ee{\end{equation}}
\def\bea{\begin{eqnarray}}
\def\eea{\end{eqnarray}}

\title{Lattice models for ballistic aggregation in one-dimension}
\shorttitle{Lattice models for ballistic aggregation}

\author{Supravat Dey$^1$ \and Dibyendu Das$^1$ \and R. Rajesh$^2$}
\shortauthor{Supravat Dey  \etal}
\institute{                    
  \inst{1} Department of Physics, Indian Institute of Technology 
Bombay, Powai, Mumbai-400076, India\\
  \inst{2} Institute of Mathematical Sciences, CIT campus, Taramani, 
Chennai-600113, India
}

\date{\today}

\abstract{
We propose two lattice models in one dimension, with stochastically 
hopping particles which aggregate on contact. The hops are guided by 
``velocity rates" which themselves evolve according to the rules of 
ballistic aggregation as in a sticky gas in continuum. Our lattice 
models have both velocity and density fields and an appropriate real 
time evolution, such that they can be compared directly with event 
driven molecular dynamics (MD) results for the sticky gas. We 
demonstrate numerically that the long time and large distance behavior 
of the lattice models are identical to that of the MD, and some exact 
results known for the sticky gas. In particular, the exactly predicted 
form of the non-Gaussian tail of the velocity distribution function is 
clearly exhibited. This correspondence of the lattice models and the 
sticky gas in continuum is nontrivial, as the latter has a deterministic 
dynamics with local kinematic constraint, in contrast with the former; 
yet the spatial velocity profiles (with shocks) of the lattice models 
and the MD have striking match.}


\begin{document}
\maketitle

\section{Introduction}

A gas of ballistically moving particles which suffer completely 
inelastic collisions amongst themselves and thereby aggregate, is well 
known as the {\it sticky} gas. The motion of the particles are 
deterministic and the collisions follow the momentum and mass 
conservation laws. The problem becomes statistical as the initial 
velocities of the particles are assumed to be random. The system 
exhibits large scale density clustering with a growing coarsening length 
scale, and appearance of shocks in spatial velocity profile. The 
connection between this problem and shock dynamics of a Burgers fluid in 
the high Reynolds number limit was established in \cite{Kida}. The 
sticky gas problem has some relevance for interstellar matter, as 
similarities between self gravitating matter and inertial Burgers fluid 
was shown in \cite{zeldovich}. In the context of terrestrial dissipative 
granular matter, a mean field analysis of the model with several scaling 
predictions was introduced in \cite{Carnevale}. The model was exactly 
solved for several statistical functions in 
\cite{FrachebourgPrl,FrachebourgPhysica}.

{\it Apriori} the sticky gas, which is a gas of particles with zero 
restitution coefficient ($r$) may appear to be rather artificial, since 
real granular gases all have $0<r<1$. In fact in reality, $r$ is also 
dependent on the relative velocity $v_{\rm rel}$ of collisional impact 
--- for the $v_{\rm rel}\rightarrow 0$, $r \rightarrow 1$ 
\cite{Raman,Brilliantov}. Very interestingly it has been shown that for 
intermediate times (but not very large \cite{Mahendraprl}) such a 
realistic granular gas with any $r$ ($0<r<1$) behaves statistically 
similarly as a sticky gas ($r=0$) \cite{Ben99,MahendraPre}. This 
remarkable universality, hitherto is only supported by molecular 
dynamics (MD) and not proved analytically. In particular, since the 
mapping between the particle-dynamics of the sticky gas and 
shock-dynamics of the Burgers fluid is known \cite{Kida} and several 
properties of the sticky gas and the Burgers equation has been derived 
analytically \cite{FrachebourgPrl,FrachebourgPhysica,Frachebourg}, one 
hopes that the speculation of the inviscid Burgers equation being the 
correct continuum limit of the finite $r$ granular gas should be 
analytically provable.

In certain non-equilibrium problems like diffusion limited reactions 
\cite{tauber,CardyNon,wijland}, a quantum field theoretic formalism has 
been develop to systematically connect the stochastic dynamics of 
appropriate lattice models to their respective continuum Langevin 
dynamical limits. Motivated by the latter works we wonder if a similar 
method can be followed for the granular gas problem discussed above. Can 
one invent suitable stochastic lattice models which may represent the 
microscopic ballistic granular gas problem? The proposal is rather 
ambitious for $r \neq 0$, but one may start with the simpler sticky gas 
(i.e. the $r=0$ case). In this paper, we propose two stochastic lattice 
models (with single and multi-particle occupancies per site 
respectively) which are shown to have the same statistical properties, 
at long times and large distances, as the sticky gas. Statistical 
distribution functions of the mass and velocity, spatial density-density 
correlation functions, and velocity shock profiles, for our lattice 
models are shown to have striking match with exact results and MD 
simulation results of the sticky gas. Of the two models we discuss 
below, the single particle occupancy model would seem a natural 
counterpart of the sticky gas. On the other hand, the second model with 
multiple site occupancy and a delayed aggregation would apriori seem a 
bit different --- although in the limit of a tuning parameter we show 
that it reduces to the single occupancy model. Explicit study of a 
multiple particle occupancy model is motivated by the fact that the 
field theoretic literature \cite{tauber,CardyNon} have generally dealt 
with multiple occupancy models. Furthermore they are also natural 
for future extension of our study to the $r\neq0$ case, where particles 
do not stick after collisions. Thus our choice of the models leave open 
the scope of coarse graining in future, and obtaining the desired 
continuum limit.

Although lattice models have been studied earlier 
\cite{Puglisi01,Puglisiepl,Puglisi02,inelasmaxwell,Nienhuis} for the 
freely cooling granular gas, to understand the velocity ordering and 
formation of spatial shocks they had many limitations. Firstly, these 
earlier models only had lattice velocities which evolved stochastically 
and they involved no actual particles hopping. As a result although key 
insights on the behavior of the velocity field could be obtained, the 
``density field" could not be tracked. Moreover the time used for 
characterizing evolution of the system was ``collision number", and 
matching to real time evolving MD involved arbitrariness. Finally 
kinematic rules had to be specified separately on the lattice models to 
avoid collisions disallowed by the ballistic dynamics. In the models 
that we present in this paper, these short-comings have been overcome. 
We have actual particles hopping and so both the velocity and mass 
density fields can be tracked. The time evolution is done using 
prescriptions of exact stochastic simulation of Master equation, such 
that it compares directly with time in MD simulations. Interestingly the 
stochastic dynamics in our models naturally take care of one type of 
kinematically forbidden collisions, but violates another type (a 
slower particle can catch up with a faster one from behind). In spite of 
this lack of strict adherence to kinematic constraints we show that the 
results of velocity shock profiles are nevertheless almost identical to 
MD for identical initial conditions. Thus we claim our lattice models to 
be minimally complex models necessary to represent the sticky gas 
system, and may hope (as discussed above) that in future they may be 
appropriately coarse-grained to derive the inviscid Burgers equation.

Below we begin by first defining the single and multiple occupancy 
lattice models SOSG and MOSG, respectively.

\section{Model I: Single occupancy sticky gas (SOSG)}

The system has massive particles on a one-dimensional periodic lattice 
of size $L$. An occupied site $i$ at time $t$ (where i can take values 
$1,2,...,L$) has two non-zero variables: mass $m_i$ and ``velocity" 
$v_i$. The magnitude of the velocity $v_i$ specifies the rate of hopping 
of the particle to its nearest neighbors, and its sign decides the 
direction of hopping. For an unoccupied site we designate $m_i = 0$ and 
$v_i = 0$. We initialize the system by having $m_i=1$ $\forall i$ and 
assigning random velocities $v_i$ drawn from an uniform box distribution 
over the range $(-1, +1)$. The system evolves in time via hopping and 
instantaneous aggregation events. Let the sites $i$ and its neighbor $j$ 
have variables $(m_i, v_i)$ and $(m_j, v_j)$ respectively at time $t$. 
The particle of the site $i$ may hop to either of its neighbors $j (= 
i+1, i-1)$ depending on $sgn(v_i)(= +1, -1)$, and with the rate $|v_i|$. 
If at time $t+\delta t$ the particle at the $i^{th}$ actually hops to 
the $j$ then the masses and the velocities change in the following way: 
$m_i\rightarrow 0$, $v_i\rightarrow 0$, $m_j\rightarrow m_i + m_j$, and 
$v_j\rightarrow (m_iv_i + m_jv_j)/(m_i+m_j)$. The latter update of $v_j$ 
follows the momentum conservation rule of particle collision with zero 
restitution coefficient.

To implement the above dynamics, in our stochastic simulation, we choose 
an event of hopping of a particle at site $i$ with probability 
$|v_i|/\Gamma$ \cite{lebowitz,gillespie76,gillespie77,gillespieRev}, 
where $\Gamma = \sum_i |v_i|$ is the sum of all hopping rates possible 
in the system at the time $t$. The enactment of the event is associated 
with a time increment $\delta t$, which we choose to be the reciprocal 
of the total rate $\Gamma$, i.e $\delta t = 1/\Gamma$. As $\Gamma$ is 
time dependent in this problem, $\delta t$ is also time dependent. We 
note that this choice of $\delta t$ is the average value of the truly 
random $\delta t$ drawn from its distribution function $\Gamma 
\exp(-\Gamma \delta t)$ 
\cite{lebowitz,gillespie76,gillespie77,gillespieRev}. Although using a 
random $\delta t$ would have been more appropriate in principle, for 
very small rate $\Gamma$ as we have at large times, sampling its full 
distribution becomes very tedious numerically. Using instead the average 
value $\delta t = 1/\Gamma$ saves numerical effort considerably, without 
compromising accuracy (as we have checked).

\section{Model II: Multiple occupancy sticky gas (MOSG)}

In this model we also have a system of massive particles on an 
one-dimensional periodic lattice of L sites, except that now multiple 
particles can occupy any site. For a site $i$ occupied by $\alpha^{th}$ 
particle, we associate two variables $m_{i,\alpha}$ and $v_{i,\alpha}$. 
If the number of particles at site $i$ is $n_i$, then $\alpha$ can take 
values $1,2,...,n_i$. As in SOSG $|v_{i,\alpha}|$ denotes the rate of 
hopping while $sgn(v_{i,\alpha})(=+1,-1)$ decides to which neighboring 
site $j=(i+1,i-1)$ the particle may hop. The system evolves via hopping 
followed by aggregation, except that in MOSG unlike SOSG, aggregation 
events are not instantaneous. At any site $i$, a pair of particles 
$\alpha$ and $\beta$ may aggregate with a rate $\lambda$. In the limit 
of $\lambda\rightarrow\infty$, aggregation events becomes instantaneous 
leading to maximum occupancy of the sites being $1$ --- thus MOSG maps 
back to SOSG. We initialize the system by having $n_i=1$ and 
$m_{i,\alpha}=1$ $\forall i$, and assigning random velocities 
$v_{i,\alpha}$ drawn from an uniform box distribution over the range 
$(-1, +1)$. After aggregation of two particles $\alpha$ and $\beta$ at 
site $i$, $m_{i,\alpha}\rightarrow m_{i,\alpha}+m_{i,\beta}$, 
$v_{i,\alpha}\rightarrow (m_{i,\alpha}v_{i,\alpha} + 
m_{i,\beta}v_{i,\beta})/(m_{i,\alpha}+m_{i,\beta})$, and $n_i\rightarrow 
n_i-1$.

As in SOSG the stochastic simulation is implemented by choosing events 
of hopping of particles and their aggregation randomly with 
probabilities $|v_{i,\alpha}|/\Gamma$ and $\lambda/\Gamma$ respectively. 
Here $\Gamma = \sum_{i,\alpha} |v_{i,\alpha}|+\sum_i 
n_i(n_i-1)\lambda/2$ is the sum of all hopping and aggregation rates 
possible in the system at the time $t$. Note that the factor 
$n_i(n_i-1)/2$ comes from the number of possible collisions among any 
pair of particles at site $i$. The increment of time between two events 
is chosen to be $\delta t=1/\Gamma$.

As noted above, both the models SOSG and MOSG are more realistic in 
having on one hand both local mass density and velocity fields, and on 
the other hand realistic time increments $\delta t$. In both these 
respects we overcome serious drawbacks of the earlier lattice models 
\cite{Puglisi01,Puglisiepl,Puglisi02,Nienhuis} of dissipative gases in 
making contact with molecular dynamics and continuum theories. At the 
same time the models being stochastic and not having strict kinematic 
constraint (as will be discussed below), leave the curious question open 
as to whether their long time and large wavelength properties would 
match with that of the deterministically evolving ballistic aggregation 
model. We now proceed to show numerically that such a non-trivial 
correspondence is indeed exhibited by both the models.

\section{Results}

For both SOSG and MOSG lattice models we use system sizes $L=50000$. As 
mentioned before the hopping rates $v_i$ (in SOSG) and $v_{i,\alpha}$ 
(in MOSG) are initially chosen to be random and drawn from the uniform 
box distribution over range (-1,1). In MOSG we have used the aggregation 
rate $\lambda=5.0$. The unit of time $t$ in our lattice simulations is 
inverse of the unit of rates --- as rates are chosen to be 
dimensionless, the time is also dimensionless. Similarly the space $x$ 
in units of the lattice spacing is taken as dimensionless.

\begin{figure}[htbp]
\includegraphics[scale=.65]{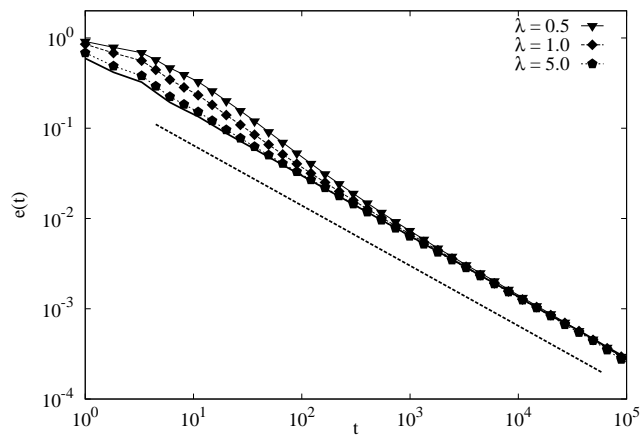}
\caption{\label{et}Log--log plot of Energy $e(t)=E(t)/E_0$ with time t, where $E(t)$ is the total energy  at time $t$ and $E_0=E(t=0)$. Data points with lines joining them are for the MOSG model with different values of $\lambda$ indicated as labels. The solid line is for the SOSG model. The dashed line has slope -2/3 and serves as a guide to eye.}
\end{figure}

The choice of the value of $\lambda=5.0$ is based on the observation of 
energy decay in MOSG for various values of $\lambda$ shown in Fig. 
\ref{et}. The scaled energy $e(t)$ for the MOSG model with different 
$\lambda$, tend to that of the SOSG model for increasing $\lambda$. For 
large $t$ both the models follow the classic energy decay $e(t)\sim 
t^{-2/3}$ \cite{Carnevale,Ben99,FrachebourgPhysica} well known for the 
ballistic sticky gas. In particular for $\lambda=5.0$ we can see from 
Fig. \ref{et} that for times $t\geq10^2$ the MOSG and SOSG curves are 
completely merge, implying that we are safely in the universal sticky 
gas regime. Interestingly the irrelevance of $\lambda$ and the 
universality of $e(t)$ at large $t$, is reminiscent of a similar 
irrelevance of the values of restitution coefficient in connection to 
energy decay of granular gases in one-dimension \cite{Ben99}.

Exact analytical results in terms of explicit functions and integrals 
for the mass and velocity distribution functions of the sticky gas are 
available \cite{FrachebourgPhysica,FrachebourgPrl}. On the other hand 
the mass density-density spatial correlation function, although not 
available as an explicit analytic form, can easily be obtained from 
event driven molecular dynamics (MD) simulation of the sticky gas 
\cite{MahendraPre}. Below we first proceed to match the numerically 
obtained mass and velocity distributions from the SOSG and MOSG models 
with the exact analytical results. Then we compare the mass density 
correlator of SOSG and MOSG models with that of MD. Finally, even the 
microscopic spatial velocity shock profile of the two lattice models are 
shown to be almost identical to that of the MD.

\subsection{Comparison of the lattice model results with the analytical results}

The exact analytical solution of the sticky gas problem 
\cite{FrachebourgPhysica,FrachebourgPrl} relies on reducing the 
calculation of the distribution functions to a summing of Brownian paths 
with parabolic constrains in momentum space. The diffusion constant $D$ 
associated with the effective Brownian motion is exactly equal to the 
inverse of the variance $\beta$ of the initial momentum distribution of 
the particles in the sticky gas\cite{FrachebourgPhysica}. The following 
distribution functions for the scaled mass $M = m/t^{2/3}$ and scaled 
velocity $V = vt^{1/3}$ were derived in \cite{FrachebourgPhysica} as 
follows:
\bea
\mu_1(M)&=&\beta M{\cal I}(M){\cal H}(M)
\label{epm}\\
\bar\mu_1(V)&=& 2 \left(\frac \beta 2\right)^{4/3} \int_0^\infty dM\,M{\cal I}(M) \times \nonumber \\
&&~~~~~~~~~~~{\cal J}(V-M/2){\cal J}(-V-M/2)
\label{epv}
\eea
where,
\bea
{\cal H}(M)&=&{1\over 2i\pi }\int_{-i\infty}^{+i\infty}
dw{{\rm e}^{-{(\frac{\beta}{2})}^{1/3} M w}
\over {\rm Ai}^2(w)} \nonumber\\
\label{hhh}
{\cal I}(M)&=&\sum_{k\geq 1} {\rm e}^{-{(\frac{\beta}{2})}^{1/3} \omega_k M}\nonumber\\
\label{im}
{\cal J}(Y)&=&{1\over 2i\pi }\int_{-i\infty}^{+i\infty}dw
{{\rm e}^{{(\frac{\beta}{2})}^{1/3}Yw}\over {\rm Ai}(w)}.
\label{calj}
\eea
In Eq. \ref{calj} Ai is the Airy function \cite{Arfken} and $-\omega_k$ 
(k=1,2,3,...) are its zeros.

The only parameter appearing Eqs. \ref{epm}, \ref{epv}, and \ref{calj} 
is $\beta$ which has to be fixed to match these formulas with our 
simulation results. In both the lattice models the initial mass of every 
particle is unity, and hence the variance of the momentum distribution 
is same as that of the velocity distribution. The variance of the 
uniform box distribution of velocities over the range (-1,1) is 
$\beta=1/3$. Using the latter value of $\beta$, we evaluate the 
integrals ${\cal {H}}$ and ${\cal {J}}$ using the ``quadgk" function 
with absolute tolerance $=10^{-14}$ in MATLAB. The evaluation of the sum 
${\cal I}(M)$ was done using our own code --- for large $M$ ($M\geq0.1$) 
we kept up to 6000 terms in the sum, while for small $M$ ($M<0.1$) we 
used the asymptotic formula ${\cal I}(M)\simeq 1/\sqrt{2\pi\beta 
M^3}$\cite{FrachebourgPhysica}.
\begin{figure}[htbp]
\includegraphics[scale=.65]{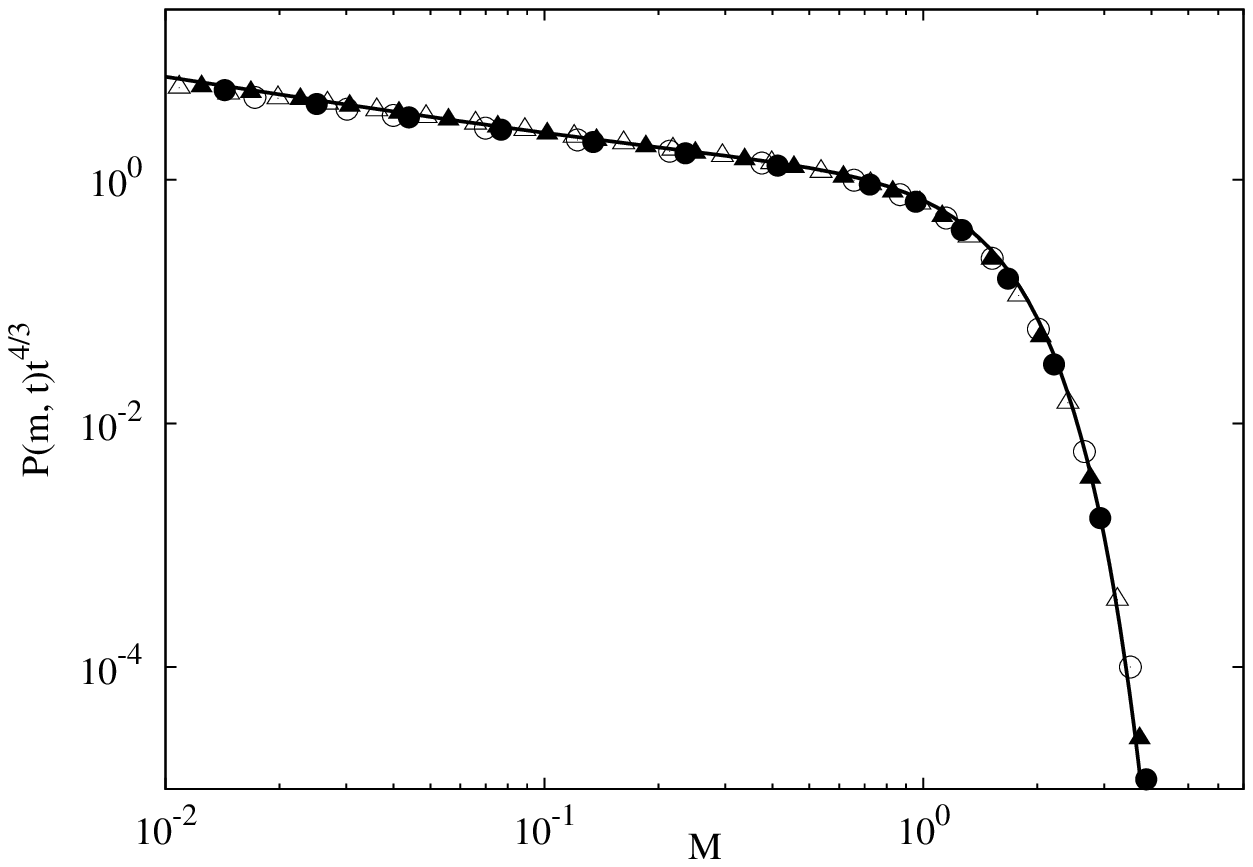}
\caption{\label{md}Log--log plot of $P(m,t)t^{4/3}$ versus $M=m/t^{2/3}$ for two different models at two different times. Circular symbols ($\fullmoon$, $\newmoon$ ) represent the SOSG data, while triangular symbols ($\triangle$, $\blacktriangle$) represent the MOSG data. Empty symbols are data for $t=8\times10^3$ and solid symbols are data for $t=16\times 10^3$. The solid line is the exact analytical formula of Eq. \ref{epm}.}
\end{figure}

The mass distribution function $P(m,t)$ is related to the scaling 
function $\mu_1(M)$ (Eq. \ref{epm}) as follows:
\be
P(m, t) = \frac{1}{t^{4/3}}\mu_1(M).
\label{pm}
\ee 
In the simulation of the SOSG model at any given time $t$, we drew the 
frequencies of the non-zero integer masses sweeping over all the sites 
of the lattice. In contrast, in the MOSG model the frequencies of the 
total mass ($\sum_{\alpha=1}^{n_i}m_{i,\alpha}$) of every occupied site 
were drawn, again sweeping over the lattice sites. Normalizing the 
frequencies by $L$ and averaging these further over several random 
initial conditions ($\sim10^4$) we finally obtained the mass 
distribution function $P(m,t)$. The latter distribution functions 
obtained for the stochastic SOSG and MOSG models at two different times 
namely $t=8\times10^3$ and $t=16\times10^3$ are scaled (Eq. \ref{pm}) 
and plotted in Fig. \ref{md}. The exact scaling function $\mu_1(M)$ 
enumerated by the method described above is plotted against the data of 
the lattice models. The spectacular match of the Monte Carlo data and 
the exact formula, without any parameter fitting, gives the first 
evidence to our claim that the lattice models SOSG and MOSG are correct 
representatives of the continuum behavior of the sticky gas.

We now proceed to look at the velocity distributions for the lattice 
models. The normalized distribution of our interest scales as follows:
\be
Q(v, t) = {t^{1/3}}f_1(v t^{1/3}),
\label{pv}
\ee 
where $f_1(V)=\bar\mu_1(V)/{\int_{-\infty}^{\infty} {\bar\mu_1(V)}}$, 
and $\bar{\mu_1}(V)$ is given by Eq. \ref{epv}. Unlike the integer 
masses $m$, the velocities are real numbers, so a numerical binning is 
required to obtained a distribution function. For SOSG, at any time $t$ 
we find the frequencies of velocities of occupied sites falling within 
coarse-grained bin widths of magnitude $0.001$. For MOSG, the 
frequencies of average velocities 
($\sum_{\alpha=1}^{n_i}v_{i,\alpha}/n_i$) of occupied sites falling 
within different bins (each of width $=0.001$) were obtained. We 
normalize these frequencies by the total number of occupied sites and 
the bin width. Since the tail of the velocity distribution is of great 
interest, to obtain high precision, we had to average over relatively 
much larger set of initial conditions in this case (namely $\sim 10^7$).
\begin{figure}[htbp]
\includegraphics[scale=.65]{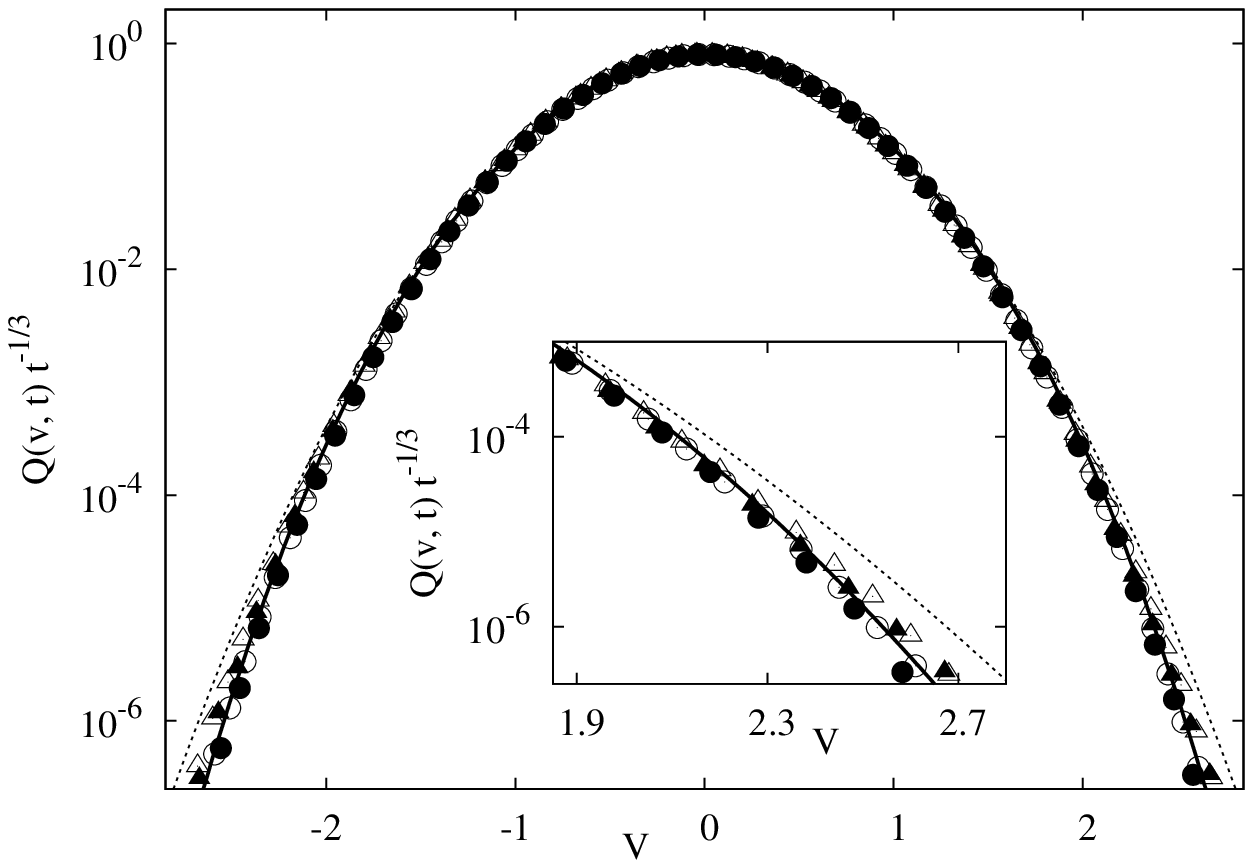}
\caption{\label{vd}
Log--linear plot of $Q(v,t)/t^{1/3}$ versus $V=vt^{1/3}$ for two 
different models at two different times. Circular symbols ($\fullmoon$, 
$\newmoon$ ) represent the SOSG data, while triangular symbols 
($\triangle$, $\blacktriangle$) represent the MOSG data. Empty symbols 
are data for $t=8\times10^3$ and solid symbols are data for $t=16\times 
10^3$. The solid line is the $f_1(V)$ using exact analytical formula for 
$\bar\mu_1(V)$(Eq. \ref{epv}). The dotted line is a Gaussian curve with 
zero mean and variance $=0.26$ which fits only the small $V$ data. 
Inset: A zoom into the tail region of $Q(v,t)/t^{1/3}$ versus $V$ (for 
positive $V$) to show clearly the deviation of the data and the exact 
formula from the Gaussian form.}
\end{figure}

The velocity distribution data obtained from the lattice models SOSG and 
MOSG at two different times namely $t=8\times10^3$ and $t=16\times10^3$ 
are scaled (Eq. \ref{pv}) and plotted in Fig. \ref{vd}. The exact curve 
of $f_1(V)$ (obtained using $\bar\mu_1(V)$ of Eq. \ref{epv}) shown with 
a solid line in Fig. \ref{vd} passes all the way through the Monte Carlo 
data. Again the spectacular match of the data and the exact result 
without any parameter fiddling is to be noted. A Gaussian curve shown in 
dotted line fits the data for small $V$, but the tails of the data 
depart from it.

The asymptotic form of the Eq. \ref{epv} for large $V$, i.e. the 
tail, is not known exactly.  In \cite{FrachebourgPhysica} the tail was 
shown to be bounded as $\bar\mu_1(V\rt\infty) \leq {\rm 
constant}~|V|\exp(-{\beta |V|^3}/{6}-({\beta/2})^{1/3}|V|\omega_1)$. The 
inset of Fig. \ref{vd} shows the Monte Carlo data near the tail; the 
deviation from Gaussianity and match with the exact Eq. \ref{epv} are 
conclusive. However, to determine the precise asymptotic decay, we will 
require data for much larger range of $V$. At this point we would like 
to recall that finding non-Gaussianity in numerical study of freely 
cooling granular gases has been an extremely challenging task in earlier 
works. For example in one-dimensional granular gas with finite 
restitution coefficient \cite{Ben99}, although the system was argued to 
approach the sticky gas limit asymptotically, the numerical data for the 
velocity distribution did not clearly show deviation from Gaussianity. 
We have checked that the computation time involved in MD is very huge to 
attain the necessary accuracy to demonstrate the deviation of the tail 
of the distribution from the Gaussian form. Interestingly in a very 
different context of quasi-elastic limit of granular gas, `stationary' 
velocity distribution with tail of the form $\exp(-c|V|^3)$ have been 
numerically demonstrated \cite{TrizacJphysA} but that does not directly 
address the concerned limit ($r=0$) of this paper. Given this background 
it is significant that we have succeeded in demonstrating the deviation 
from Gaussianity of the tail of the velocity distribution with 
comparative ease in the Monte Carlo simulations of the lattice models 
SOSG and MOSG.

\subsection{Comparison of the lattice model results with MD results}

Event driven MD of ballistically moving sticky particles in continuum is 
expected to yield numerically exact results, if done with sufficient 
accuracy. The details of such simulation can be found in a recent 
publication \cite{MahendraPre}. Here we use event driven MD to study the 
spatial mass density-density correlation function and spatial velocity 
profile, compare these with that obtained from Monte-Carlo simulation of 
the SOSG and MOSG models.

The local mass density function is defined in the following fashion. For 
the SOSG and MOSG models a local site mass density $m_i$ is defined as 
the total mass at site $i$. For the MD simulation the whole continuum 
space of length $L$ is divided into $L$ boxes and the sum total of 
masses in the $i^{th}$ box denotes the mass density $m_i$ of the box 
\cite{MahendraPre}. The density-density correlation function 
$C_{mm}(x,t)=\langle m_i(t)m_{i+x}(t)\rangle$. Here, the angular bracket 
$\langle.\rangle$ represents average over both space ($i$) and initial 
velocities, in all the three systems. In MD simulations with $L$ 
particles, the system is initialized to have random particle velocities 
drawn from an uniform box distribution over $(-1, 1)$, and unit particle 
masses kept at unit separations. In MD too we use $L=50000$. Thus the 
initial conditions are same for MD and the two lattice models, although 
their subsequent dynamics are very different.
\begin{figure}[htbp]
\includegraphics[scale=.65]{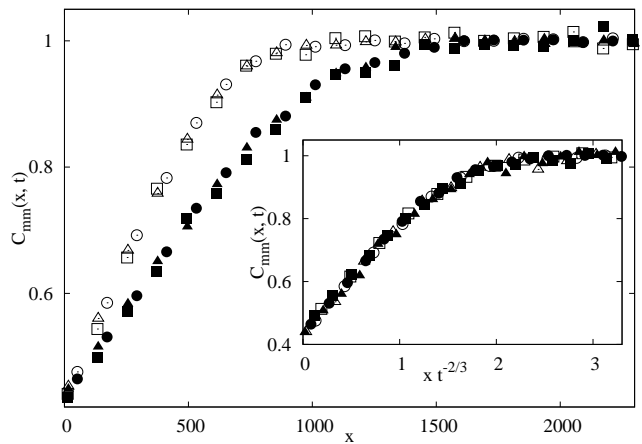}
\caption{\label{dc} Plot of $C_{mm}(x, t)$  versus $x$ for the MD simulations and the two different lattice models, at two different times. Empty symbols are data for $t=8\times10^3$ and solid symbols are data for $t=16\times 10^3$. Circular symbols ($\fullmoon$, $\newmoon$ ) represent the SOSG data, triangular symbols ($\triangle$, $\blacktriangle$) represent the MOSG data, and the square symbols ($\square$, $\blacksquare$) represent the MD data. Inset shows plot of $C_{mm(x,t)}$ versus $x/{\cal L}(t)$ for the six curves in the main figure --- the data collapse according to Eq. \ref{dceq}.}
\end{figure}

The mass density-density correlation functions are shown in Fig. 
\ref{dc} at two different times. At both the time instants the data of 
the lattice models and the MD simulations fall perfectly on top of each 
other, without any parameter fixing. The profile of the correlation 
function is characteristic of the sticky gas and has been discussed in 
\cite{MahendraPre}. As the system coarsens in time, $C_{mm}(x,t)$ is 
expected to obey the scaling hypothesis \cite{porod} when space is 
scaled with the coarsening length ${\cal L}(t)\sim t^{2/3}$:
\be
C_{mm}(x, t) = g\left(\frac{x}{{\cal L}(t)}\right).
\label{dceq}
\ee
As shown in the inset of the Fig. \ref{dc}, there is perfect data 
collapse following the above equation.
\begin{figure}[htbp]
\includegraphics[scale=.65]{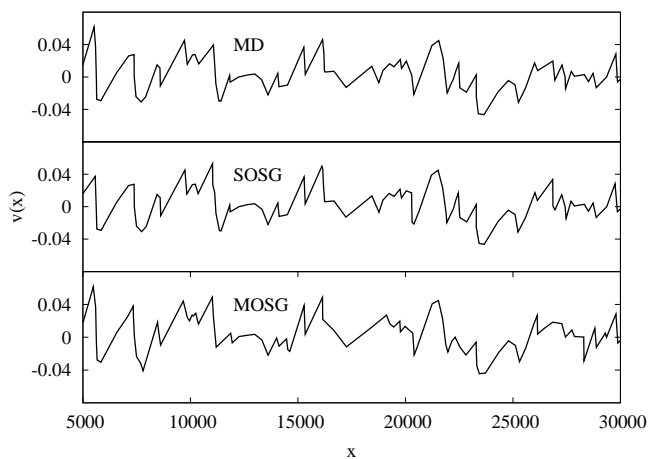}
\caption{\label{shocks} Velocity profiles $v(x)$ plotted against space 
$x$ (which is continuous for MD and assumes integer values for the SOSG 
and MOSG models). The profiles of $v(x)$ at time $t=10^4$ look almost 
identical in the three cases, for the same initial realization of random 
velocities.}
\end{figure}

Unlike spatial correlation functions which capture ordering and 
structure formation at the macroscopic level, microscopic information is 
captured by the spatial velocity profile of a sticky gas. The velocity 
profile $v(x)$ exhibit shocks joined by approximate linear curves 
\cite{Kida,Ben99,Nienhuis}. In Fig. \ref{shocks} we show 
particle-velocity $v(x)$ versus $x$ for the MD simulation with visible 
shocks (where velocity jumps suddenly from positive to negative) as 
expected. It is remarkable that for the SOSG and the MOSG lattice models 
the velocity profile of occupied site at the same time $t$, are almost 
identical as the MD (Fig. \ref{shocks}) provided their initial 
conditions are identical. Although in an earlier work such a 
correspondence was demonstrated \cite{Nienhuis}, there the shocks in the 
velocity profile had an inverse correspondence to the shocks in MD --- 
lattice shocks were located at positions where MD profile was gradual 
and vice versa; this was a consequence of the fact that there were no 
actual motion of particles along the lattice. Contrary to that since we 
have actual particles hopping on lattice, we have natural and direct 
correspondence of shock profiles with the MD (Fig. \ref{shocks}). 
Moreover it was emphasized in \cite{Nienhuis} that local kinematic 
constraints had to be separately imposed on the lattice dynamics to 
ensure that collisions disallowed by the ballistic dynamics do not 
occur, and this was argued to be a key factor for shock formation. The 
kinematic constraint states that if $v_R - v_L > 0$ (where $v_L$ and 
$v_R$ denote velocities of a pair of left and right particles, 
respectively) collisions won't occur. The collisions disallowed by the 
latter rule can be classified into two types: (a) $v_R>0$ and $v_L<0$, 
(b) $v_R>v_L$ with same sign for both. In our simulation since hopping 
in the opposite direction of rate $v$ is disallowed, case (a) is 
automatically taken care of --- this we would claim is indeed the key 
condition for shock formation. On the other hand, due to the stochastic 
nature of hopping in our models, constraint (b) is sometimes violated (a 
slower particle may catch up with a faster one) --- yet the latter 
violation has no discernible effect (Fig. \ref{shocks}) on the 
microscopic shock profile. Hence we conclude that constraint (b) is not 
an essential constraint.

\section{Discussion}

We have shown that two lattice models with single and multiple particles 
site occupancy and stochastic dynamics, can exactly reproduce the long 
time and large space behavior of the ballistic sticky gas system. In 
particular, the mass and velocity distribution functions are identical 
to the exact analytically known distributions of the sticky gas. The 
density-density spatial correlation functions and local microscopic 
velocity shock profiles of the lattice models and MD simulations also 
have a perfect match.

As we have noted that the current lattice models differ from earlier 
lattice models in having a genuine density field besides the velocity 
field. The time update is done following a prescription suited for real 
continuous time evolution, allowing us to directly compare our results 
with exact analytical and MD simulation results. The lattice dynamics 
violates the strict conditions of kinematic constraint, but only mildly, 
so that there is no observable effect over slightly coarse-grained 
spatial scale (see Fig. \ref{shocks}). 

Based on the results that we have found, a curious possibility has been 
opened up to derive the continuum behavior of the granular gases in 
future. We hope that the two lattice models studied here have all the 
necessary ingredients sufficient to capture the correct continuum 
behavior of the sticky gas; the models should lead to the inviscid 
Burgers equation if they can be successfully coarse grained by a 
suitable technique.

\end{document}